\documentclass[twocolumn,showpacs,preprintnumbers,amsmath,amssymb,pra]{revtex4}

\usepackage{graphics}
\usepackage{epsfig}
\usepackage{bm}

\def\beq{\begin{equation}}
\def\eeq{\end{equation}}

\def\reff#1{(\ref{#1})}

\def\subsc#1{{\mbox{\rm\scriptsize #1}}}

\def\rhocrit{\rho_\mathrm{c}}
\def\rhorat{\rho/\rho_\mathrm{c}}
\def\Up{U_\mathrm{p}}

\def\rbar{\bar{r}}

\def\Etot{E_\mathrm{tot}}

\def\Wcmcm{\mbox{\rm Wcm$^{-2}$}}
\def\abl#1#2{\frac{\mbox{\rm d} #1}{\mbox{\rm d} #2}}

\def\N3d{N_\subsc{3D}}

\def\omegaMie{\omega_\mathrm{Mie}}

\def\omegalaser{\omega_\mathrm{l}}

\def\omegalasersq{\omega_\mathrm{l}^2}

\def\vekt#1{\bm{#1}}

\def\vektr{\vekt{r}}

\def\vektE{\vekt{E}}

\def\vektnabla{\vekt{\nabla}}

\def\vektEsc{\vektE_\mathrm{sc}}

\def\Edach{E_0}

\def\Ehat{\Edach}

\def\diff{\,\mbox{\rm d}}

\begin{document}
\title{
Collisionless energy absorption in the short-pulse intense laser-cluster
interaction
}
\date{\today}
\author{M.\ Kundu and D.\ Bauer}
\affiliation{Max-Planck-Institut f\"ur Kernphysik, Postfach 103980,
69029 Heidelberg, Germany}
\date{\today}

\begin{abstract}
In a previous Letter [Phys.\ Rev.\ Lett.\ {\bf 96}, 123401 (2006)] we have shown by means of three-dimensional particle-in-cell simulations and a simple rigid-sphere model that nonlinear resonance absorption is the dominant  collisionless absorption mechanism in the intense, short-pulse laser cluster interaction. In this paper we present a more detailed account of the matter. In particular we show that the absorption efficiency is almost independent of the laser polarization. In the rigid-sphere model, the absorbed energy  increases by many orders of magnitude at a certain threshold laser intensity. The particle-in-cell results display maximum fractional absorption around the same intensity. We calculate the threshold intensity and show that it is underestimated by the common over-barrier  ionization estimate.  
\end{abstract}


\maketitle

\section{introduction}\label{sec1}
Intense laser-matter interaction provides a route to 
create energetic particles (e.g., electrons, ions, and photons) 
using  table-top equipment. 
Clusters, possessing the transparency of gas targets and the 
high charge density of
solid targets, 
proved to be very efficient absorbers of laser light. 
Their small size, compared to laser wavelength and skin depth, 
avoids reflection of the laser beam at the cluster
surface as well as the loss of hot electrons into the cold bulk. 
In fact, almost 100\% absorption of the laser
light was reported in experiments with rare gas clusters
\cite{ditm97}. Useful reviews on the subject are Refs.~\cite{posthumus, saal06, kra02}.

Upon irradiation of the rare gas clusters by intense laser light, electrons
first absorb energy and leave their  ``parent'' ions. This is known as
inner ionization, meaning that the electrons are still bound to the
cluster but not necessarily to their ``parent'' ions.
 The total electric field (i.e., laser plus space
charge field) inside the cluster leads to inner ionization up to
high charge states not possible with the laser field alone
(ionization ignition \cite{rose97,bauer03}). 
As the laser intensity during the pulse increases, these electrons absorb energy
from the laser field and may leave the cluster leading to the
positive charging of the cluster known as outer ionization. Thus
outer ionization  leads to a non-neutral plasma.
With the increasing outer ionization, the restoring
force of the ions counteracts ionization ignition so that the latter stops at
some point. The net positive charge left behind
finally explodes due to the Coulomb repulsion and hydrodynamic pressure, leading to the conversion of electron energy
into ion energy. Typically MeV ions and keV electrons \cite{ditmNature,kum03,spring03} are measured in experiments. 

It is clear from the described scenario that the understanding of the relevant
    	mechanism(s) of laser energy absorption 
    	leading to the heating of cluster electrons and outer ionization is of great importance for the development of a complete theoretical description. 

Laser energy absorption by electrons proceeds either through 
resonances (linear or
nonlinear)  or through non-adiabaticities (all possible types of
collisions). All of these processes lead to dephasing of the current density with respect to the laser field, which, according to Poynting's theorem, is a prerequisite for absorption.
Collisional absorption via collisions of electrons with ions are of minor
importance at near infrared wavelengths $\simeq 800$\,nm or greater
\cite{ishi00,megi03,crist04,bauer04} whereas it is the dominant absorption
mechanism at short wavelengths \cite{bauer04,sied04,crist05}, 
not studied in this paper.   The finite size of the clusters suggests that ``collisions with the cluster boundary'' may be responsible for the energy absorption. However, this viewpoint is misleading, as will be shown in Sec.~\ref{sec2b}. 

During the expansion of the ionic core, 
the decreasing charge density $\rho(t)$
leads to the decrease of the Mie plasma frequency,
$\omegaMie(t)
\equiv \sqrt{4\pi\rho(t)/3}$ (atomic units $\hbar=m=-e=4\pi\epsilon_0=1$ are used unless noted otherwise).
For very short near infrared laser pulses $\omegaMie(t)$
cannot meet the linear resonance
\beq \omegaMie(t)=\omegalaser \label{lr}, \eeq
unless the cluster has sufficiently expanded
(typically after a few hundred femtoseconds). 
Linear resonance \reff{lr}, well understood in theory, 
experiments, and simulations 
\cite{ditm96,doepp05,koell99,zam04,last99,saal03,fenn04,sied05,mart05}, 
is thus ruled out for very short pulses or during the
early cycles of a long pulse 
laser-cluster interaction 
where ion motion is negligible. In this case, nonlinear resonance 
(NLR), whose origin lies in the 
anharmonicity of the cluster potential, turns out to be the dominant collisionless absorption mechanism. In fact, for very short linearly
polarized (LP) laser pulses, it was clearly shown \cite{kundu06}
that essentially {\em all} electrons that contribute to
outer ionization pass through the NLR,
which was unequivocally identified as {\em the} collisionless
absorption mechanism in the absence of linear resonance.
The eigenfrequency $\omega[\hat{\vektr}(t)]$ of 
a (laser-) driven oscillator
in an anharmonic potential, being dependent on its excursion amplitude 
$\hat{\vektr}(t)$ 
(or the energy), may dynamically meet the NLR 
\beq \omega[\hat{\vektr}(t)] =
\omegalaser . \label{nlr} \eeq 
Due to the many body nature of the interaction, 
the identification, the separation, 
and the interpretation of the absorption mechanisms in molecular
dynamics or particle-in-cell (PIC)
simulations are often difficult. 
Recently, a method of identification of the NLR in many-body 
simulations
of rare gas clusters has been proposed \cite{kundu06}. The possible 
importance of NLR was also 
mentioned or discussed 
previously \cite{tagu04,mulser05,baumu05,korn05}. 
The rigid sphere model (RSM) \cite{mulser05,parks01,fomi03} where
electrons and ions are modeled by homogeneously charged rigid
spheres oscillating against each other is clearly an oversimplification of a real many-particle system such as a cluster. However, it proves useful for estimating the order of magnitude of the absorbed energy as well as for the calculation of the laser intensity where energy absorption is most efficient, as will be shown in the present work. Moreover, it provides physical understanding and clearly displays NLR \cite{mulser05,kundu06}.
 
The heating of cluster electrons in circularly polarized laser fields has not yet received much attention, at least theoretically.
Experiments with rare-gas clusters show
almost no effect on the x-ray emission 
\cite{lin04, kum01, tev} and ion energy distribution \cite{kris04} 
when laser light of different
ellipticity is used. Theoretically, circular polarization is particularly interesting because the above mentioned ``collisions with the cluster boundary'' are strongly suppressed in this case. Hence one may expect energy absorption being less efficient.
NLR, on the other hand, occurs in both cases, and, in fact, the energy absorption turns out to be equally efficient. Moreover, the absorbed energy compares well with the RSM predictions.

The outline of the present paper is as follows: in Sec.~\ref{sec2a} we briefly review the NLR and the RSM. In Sec.~\ref{sec2b} the RSM is extended to circular polarization, where NLR is observed as well. In Sec.~\ref{sec2c} the RSM threshold intensities are calculated.  Section \ref{sec3} is devoted to the PIC \cite{birdsall}  results for both linear and circular polarization. Finally, we summarize our results in Sec.~\ref{sec4}.

Throughout this paper we use $n = 8$-cycle laser pulses of near
infrared wavelength $\lambda = 1056\,$nm and a fixed cluster radius
$R = 3.2\,$nm unless stated otherwise. NLR is a robust phenomenon that---qualitatively---is insensitive to cluster and laser parameters.   

\section{Nonlinear resonance in the rigid sphere model (RSM)}\label{sec2}
The ion motion can be neglected in the study of energy absorption in very short laser pulses. Thus the ions just form a static, positively charged background of spherical shape.  
For not too high 
laser intensity,  the collective motion of the electrons can also be approximated 
by a homogeneous, rigid sphere of negative charge. In the simplest case the radii of ion and electron sphere are assumed to be equal.   
In a more realistic model the electron cloud expands \cite{mulser05}.
However, the method of identification of the NLR used 
in this paper is independent of the degree of electronic expansion.
The center of mass of the electron-ion system is, in good approximation,  
located at the
center of the ion sphere.
In an oscillating laser field, the homogeneously charged electron sphere
oscillates back and forth  against the positively charged ion sphere. 
\subsection{NLR in a linearly polarized laser field}\label{sec2a}
The equation of motion of the electron
center of mass in a LP laser field, polarized 
along $x$, can be written as
\beq
\abl{^2\rbar}{\tau^2} + \frac{\rbar}{r} g(r) = -
\frac{E_\mathrm{l}(\tau)}{R\omegalaser^2},\label{rsm_eom} \eeq 
where $\rbar=x /R$ is the excursion of the
electron sphere, normalized to the cluster radius $R$, $r=\vert\rbar\vert$,  $\tau=\omegalaser t$ is the normalized time in units of the laser period, and the dimensionless electrostatic
restoring force $g(r)$ is
\beq
g(r) = \left(\frac{\omegaMie}{\omegalaser}\right)^2 \times
\left\{\begin{array}{ll}
r-\frac{9r^2}{16}+\frac{r^4}{32} \;\; & 0\leq r\leq 2 \\
\frac{1}{r^2} \;\; & r\geq 2 \\
				   \end{array}\right. .
				   \label{gr}
\eeq
The first term in the upper line of \reff{gr} is the linear
force when the displacement of the electron sphere is small, 
the next two terms are the nonlinear terms which appear due to
the partial overlap of the electron cloud with the ion cloud.  
The term $r^{-2}$ in the lower line of \reff{gr} 
is the Coulomb force between the separated electron sphere and ion sphere. 
The quantity on the right hand side of \reff{rsm_eom} is 
the normalized driver strength, which is the quiver amplitude 
$E_\mathrm{l}({\tau})/\omegalaser^2$ of a free 
electron  in the laser field $E_\mathrm{l}({\tau})$  normalized to the cluster radius $R$.
\begin{figure}
\includegraphics[width=0.5\textwidth]{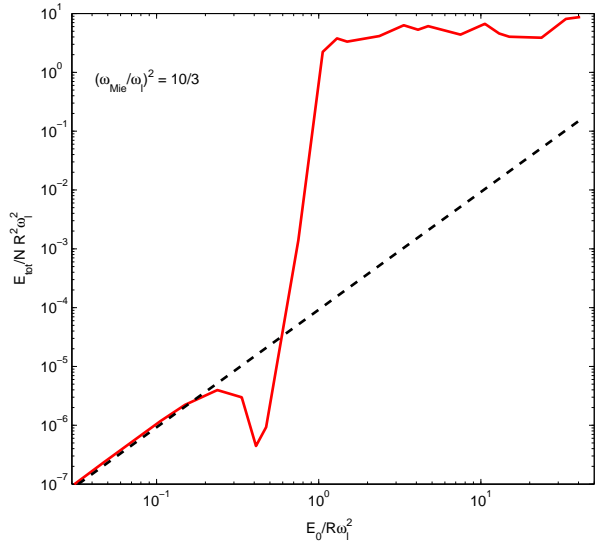}
\caption{(color online). Laser energy absorption vs driver amplitude in 
the rigid sphere-model for 
$(\omegaMie/\omegalaser)^2=10/3$,
a $n=8$ cycle laser field
$E_\mathrm{l}(\tau)=\Ehat\sin^2(\tau/2n)\cos(\tau)$, and cluster radius 
$R = 3.2$~nm. Within a narrow field strength interval (here $\simeq 0.5$--$1$) the  absorbed energy per particle  (solid line) increases by many orders of magnitude. The dashed line 
represents the absorbed energy \reff{introabseng} by a purely harmonic  oscillator driven by 
the same laser field.
\label{rsmI}}
\end{figure}
Earlier work \cite{mulser05} showed that absorption of laser energy in
the RSM is characterized by a threshold driver strength below which
absorption is negligible (harmonic regime) and above which
absorption is almost constant. 
Figure \ref{rsmI} shows this
threshold behavior for 
$(\omegaMie/\omegalaser)^2=10/3$ and
a $n=8$-cycle $\sin^2$-pulse
$E_\mathrm{l}(\tau)=\Ehat\sin^2(\tau/2n)\cos(\tau)$ for $0<\tau<2n\pi$. 
The dashed line is the absorption corresponding to a driven,   
purely harmonic oscillator
\beq
\abl{^2\rbar}{\tau^2} + 
\left(\frac{\omegaMie}{\omegalaser}\right)^2 \rbar = -
\frac{E_\mathrm{l}(\tau)}{R\omegalaser^2} 
\label{harmonicRSM}
\eeq 
in the laser field 
$E_\mathrm{l}(\tau)$. 
The laser energy absorbed by a single electron 
in a $n$-cycle laser pulse of period $T$ is 
\beq 
\frac{\Etot}{N} 
= - \int_0^{n T} \vekt v(t)\cdot \vekt E_\mathrm{l}(t)\, dt. 
\label{introdephase}
\eeq
Solving \reff{harmonicRSM} analytically for the velocity
$v(t)$ and integrating 
\reff{introdephase} one finds for the absorbed energy per electron \cite{mulser05}
\beq 
\frac{\Etot}{N} 
\simeq \frac{\omegalaser^4 \omegaMie^2
(\omegaMie^2 + 3 \omegalaser^2)^2}{4 n^4 (\omegaMie^2 - \omegalaser^2)^6}
\left[1 - \cos\left(\omegaMie n T\right)\right] E_0^2
\label{introabseng}
\eeq
where
$[\omegaMie^2 - (1 + 1/n)^2 \omegalaser^2]^2
[\omegaMie^2 - (1 - 1/n)^2 \omegalaser^2]^2
\simeq (\omegaMie^2 - \omegalaser^2)^4 $ (for sufficiently large $n$) was used.
The analytical estimate \reff{introabseng}
is plotted  in 
Fig.~\ref{rsmI} together with the 
absorbed energy obtained from 
the numerical  solution of
\reff{rsm_eom}. 
One sees that the absorbed energy jumps by many orders of magnitude to a 
higher value after crossing a threshold driver strength.
The higher the cluster
charge density is, the higher is this threshold driver strength.
The rigid sphere model shows this behavior of efficient absorption above the threshold driver strength at all 
cluster charge densities independent of the linear resonance condition
$\rho = 3 \rhocrit$, contrary to the nanoplasma model \cite{ditm96}. 
Since the rigid sphere model does not necessarily 
requires expansion of the cluster
for the efficient absorption of laser energy, it permits us to 
understand the behavior of energy absorption and the underlying
mechanism for very short laser pulses.

Equation \reff{rsm_eom}
can be formally rewritten as \beq
\abl{^2\rbar}{\tau^2}+\left[\frac{\omega_\mathrm{eff}[\hat{r}(\tau)]}{\omegalaser}\right]^2
\rbar = - \frac{E_\mathrm{l}(\tau)}{R\omegalaser^2}. \label{nlho}
\eeq 
Equation \reff{nlho} yields the instantaneous, scaled effective frequency squared
\beq \left[\frac{\omega_\mathrm{eff}(\tau)}{\omegalaser}\right]^2=
\frac{ - \frac{E_\mathrm{l}(\tau)}{R\omegalaser^2} - \ddot{\rbar}(\tau)}
{\rbar(\tau)} = \frac{g[r(\tau)]}{r(\tau)}\label{efffrequ}
\eeq which passes through unity at the NLR  \reff{nlr}. The right hand side of \reff{efffrequ} is the restoring force divided by the
excursion of the electronic cloud, which in the case of harmonic motion would be the square of the characteristic frequency.
\begin{figure}
\includegraphics[width=0.5\textwidth]{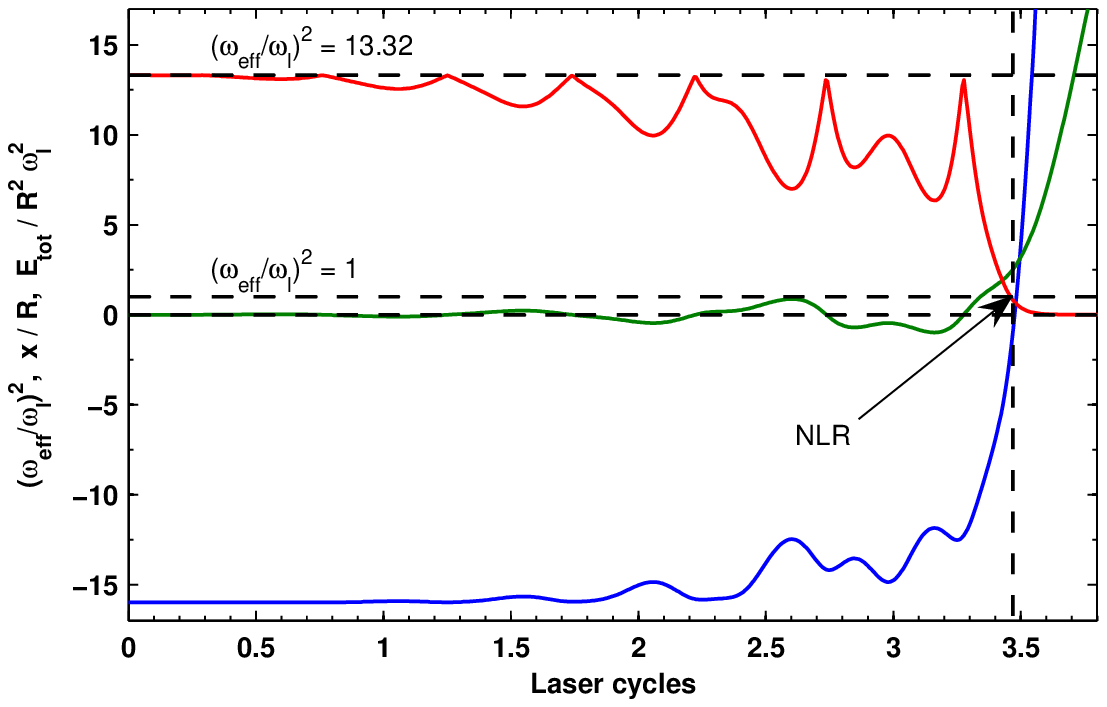}
\caption{(color online). Typical behavior of $\left({\omega_\mathrm{eff}(\tau)}/{\omegalaser}\right)^2$ (red, upper solid line) vs  laser cycles
above the threshold driver strength for 
$(\omegaMie/\omegalaser)^2=40/3$. 
Here we take $\Ehat/R\omegalaser^2\simeq 7.48$ corresponding to a laser intensity
$\simeq 2.5\times 10^{16} \mathrm{W/cm^2}$,
a $n=8$-cycle $\sin^2$-pulse $E_\mathrm{l}(\tau)=\Ehat\sin^2(\tau/2n)\cos(\tau)$
of wavelength $\lambda = 1056\,$nm, and a cluster radius $R = 3.2\,$nm.
Excursion $x/R$ (green, middle solid line) and  energy of the electron sphere $\Etot/R^2\omegalaser^2$ (blue, lower solid line) are included in the plot. Outer ionization (i.e., $\Etot/R^2\omegalaser^2\geq 0$) and occurrence of NLR $ \left[\omega_\mathrm{eff}(\tau)/\omegalaser\right]^2=1$ always 
coincide (dashed vertical line).
\label{rsmII}}
\end{figure}
Figure \ref{rsmII} shows a typical example of the temporal behavior of
$ \left[\omega_\mathrm{eff}(\tau)/\omegalaser\right]^2$ 
above the threshold
driver strength 
for $(\omegaMie/\omegalaser)^2=40/3$. 
Since $(\omegaMie/\omegalaser)^2=13.32$,
$ \left[\omega_\mathrm{eff}(\tau)/\omegalaser\right]^2$ starts at this value 
and drops with increasing driver strength. It passes through unity at the time indicated by the vertical line, and it is exactly at that time where the electron sphere becomes
free (outer ionization). This incidence is clearly visible from the energy 
of the electron sphere, which passes through zero and the excursion  as well.
Outer ionization and occurrence of NLR happens for all driver strengths above
the threshold whereas the resonance is never met below the
threshold.
Since the amplitude of the excursion of the electronic sphere
depends upon the driver strength, the excursion amplitude should
also be large enough so that the NLR is passed.
The decrease in the effective frequency with the
increase of the amplitude of excursion of the electronic sphere
in the force field \reff{rsm_eom} can be understood by
analyzing its motion in the corresponding anharmonic potential 
\beq V(r) = {\omegaMie}^2 R^2 \times \left\{ \begin{array}{ll}
\frac{r^2}{2}-\frac{3 r^3}{16}+\frac{r^5}{160}  \;\; & 0\leq r\leq 2 \\
 \frac{6}{5} - \frac{1}{r} \;\; & r\geq 2. \\
\end{array}\right. \label{rsm_pot}\eeq
The period $T$ of oscillation of the
 electronic sphere in the potential $V(r)$ can be approximated
 by a perturbation series \cite{kotkin} as
\begin{equation}
    T = \frac{1}{\sqrt{2}}\sum_{k = 0}^{\infty}
    \frac{(-1)^k}{k!}\frac{\partial^k}{\partial E_\mathrm{tot}^k}
    \oint \frac{[\delta V(r)]^k\, \diff r}{\sqrt{E_\mathrm{tot} - V_0(r)}}.
    \label{timeperiod}
\end{equation}
Here, $E_\mathrm{tot}$ is the total energy of the electronic sphere
in the ionic field, $V_0(r) = \omega_\mathrm{Mie}^2 R^2 r^2/2 $ is
the harmonic oscillator potential and $\delta V(r) = \alpha
R^3 r^3/3 + \beta R^5 r^5/5$ is the perturbation to the potential
 with $\alpha = - 9\omega_\mathrm{Mie}^2/16 R$ and $\beta =
\omega_\mathrm{Mie}^2/32 R^3$. 
The effective frequency is then $\omega_\mathrm{eff} = 2\pi/T$.
For the excursion $r <2$ and cluster
radius $R = 3.2\,$nm ($\simeq  60.4$\,a.u.) we can consider
$\vert\beta/\alpha\vert \ll 1$ and the approximate potential $\delta V(r)
\simeq \alpha R^3 r^3/3$. 
Corrections up to $k=6$  yield
$T = T_0 + T_1 + \cdots T_6$. The unperturbed period is
$T_0 = 2\pi/\omegaMie$ and the successive corrections are
$  T_1 = c_1/(3 \omegaMie^4) $,
$  T_2 = c_2/(3 \omegaMie^7)$,
$  T_3 = c_3/ (9 \omegaMie^{10}) $,
$  T_4 = c_4/ (72 \omegaMie^{13}) $,
$  T_5~= c_5/ (54 \omegaMie^{16}) $,
$  T_6 = c_6/ (3^6 \omegaMie^{19}) $ with
$  c_1 \simeq - {8 \alpha (2 E_\mathrm{tot})^{1/2}} $,
$  c_2 \simeq {5 \pi \alpha^2 E_\mathrm{tot}}$,
$  c_3 \simeq - {28.45 \alpha^3 (2 E_\mathrm{tot})^{3/2}} $,
$  c_4 \simeq {385 \pi \alpha^4 E_\mathrm{tot}^2} $,
$  c_5 \simeq - {318.6 \alpha^5 (2 E_\mathrm{tot})^{5/2}} $ and
$  c_6 \simeq {97.89 \times 2^9 \alpha^6 E_\mathrm{tot}^3}$,
respectively.

Neglecting the higher order term of $g(r)$ for $r<2$, 
a simpler approximation to the effective frequency can be derived from
\reff{efffrequ} by replacing the excursion $r$ with the excursion amplitude $\hat{r}$:
\beq \omega_\mathrm{eff}(\hat{r}) \simeq \omegaMie
\left(1 - \displaystyle {9 \hat{r}}/{16}\right)^{1/2} \simeq \omegaMie
\left(1 - \displaystyle {9 \hat{r}}/{32}\right) .
\label{rsm_freq2}
\eeq
\begin{figure}
\includegraphics[width=0.5\textwidth]{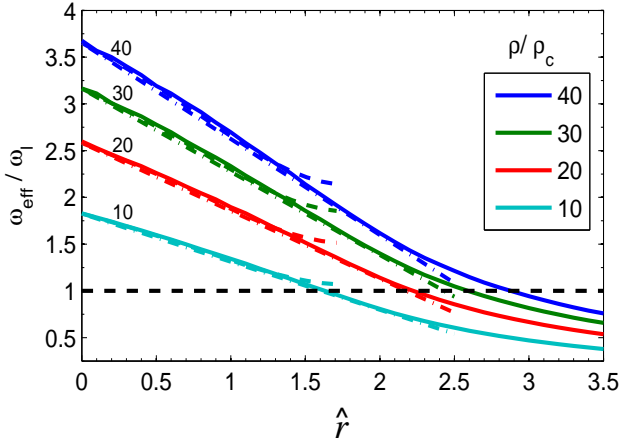}
\caption{(color online). Effective frequency 
${\omega_\mathrm{eff}}/{\omegalaser}$ vs 
the excursion amplitude $\hat{r}$ of the electronic sphere 
for different cluster charge densities
$\rho/\rhocrit = 10$--$40$.
The solid lines are computed from the numerical solution of
\reff{rsm_eom}. The dashed lines are the corresponding analytical
approximations from \reff{timeperiod} for $k$ up to 6, and the dashed-dotted
lines are from \reff{rsm_freq2}. For the charge density $\rho/\rhocrit = 40$
one expects NLR (horizontal dashed line) to occur at the excursion $\hat{r}\simeq 2.88$.
\label{rsmIII}}
\end{figure}
Figure \ref{rsmIII} shows the effective frequency vs
the excursion amplitude $\hat{r}$ of the electronic sphere in the potential
\reff{rsm_pot} for various cluster charge densities
$\rho/\rho_\mathrm{c} = 10$--$40$ as calculated from the numerical solution
of \reff{rsm_eom} together with the approximations 
\reff{timeperiod}  and \reff{rsm_freq2}. The effective frequency as calculated from
\reff{timeperiod} shows good agreement below the excursion $\hat{r}<
1.5$ and low charge densities (e.g., $\rho/\rho_\mathrm{c} <10$). For higher
charge densities more corrections [large number of $k$ values in
\reff{timeperiod}] are needed, which are very much cumbersome to calculate.
Although \reff{rsm_freq2} fits well with the numerical solution,
$\omega_\mathrm{eff}$ becomes negative when $\hat{r} > 32/9$. However, the
results using \reff{rsm_freq2} agree with the exact ones in Fig.~\ref{rsmIII} up to $\rho/\rho_\mathrm{c} = 30$  for the excursions of interest, i.e., up to the point where the  electronic sphere undergoes NLR. 
The variation of
frequency with the excursion amplitude as shown in Fig.~\ref{rsmIII} explains
why a threshold driver strength is required for an appreciable
laser energy absorption as well as for 
the crossing of the NLR in the Fig.~\ref{rsmII}: only a driver exceeding a certain threshold field strength will lead to excursions compatible with the NLR condition  $\omega_\mathrm{eff}=\omega_\mathrm{l}$.
\subsection{NLR in a circularly polarized laser field}\label{sec2b}
Clusters in a circularly polarized (CP) laser field received less attention
 in the literature. It is not known {\em a priori} how the outer ionization and
energy absorption by clusters depend on the laser polarization. In laser-atom interaction the laser polarization has dramatic effects: since in CP the free electrons do not return to their parent atom all the atomic effects relying on rescattering such as high-order harmonic generation, high-order above-threshold ionization, and nonsequential ionization are strongly suppressed. 
In the context of clusters, the study of the absorption efficiency as a function of the laser polarization can help to discriminate among different absorption mechanisms. For instance, 
if laser energy absorption was indeed due to ``collisions with the cluster boundary'' it would be suppressed in CP because the electrons mainly swirl around parallel to the ``cluster boundary'' rather than crossing (and hence colliding) with it. However, as we will show, the absorption of laser energy is largely independent of the laser polarization, thus ruling out ``collisions with the cluster boundary'' as a meaningful absorption mechanism.
\subsubsection{Two-dimensional rigid sphere model}
Let us first extend the RSM to CP. In a CP laser field with electric field components in
$x$- and $y$-direction,
the equation of motion for the electronic sphere
in the rigid sphere approximation of a cluster
can be written as
\beq\left\{ \begin{array}{c} \ddot{r}_x  \\
\ddot{r}_y
\end{array}\right\}
+ \frac{g(r)}{r}\left\{ \begin{array}{c}\displaystyle r_x  \\ \displaystyle r_y  \end{array}\right\} =
-\frac{ 1}{R\omegalaser^2}
\left\{ \begin{array}{c} E_\mathrm{l}^x(\tau) \\ E_\mathrm{l}^y(\tau)
\end{array}\right\}. \label{rsm_eomCPL}\eeq
Here, $r_x = x/R$, $r_y = y/R$,  $r=\sqrt{r_x^2 + r_y^2}$, and
\begin{eqnarray}
E_\mathrm{l}^x(\tau)&=&\frac{\Ehat}{\sqrt{2}}\sin^2(\tau/2n)\cos(\tau), \label{Ex} \\
E_\mathrm{l}^y(\tau)&=&\frac{\Ehat}{\sqrt{2}}\sin^2(\tau/2n)\sin(\tau). \label{Ey}
\end{eqnarray} 
One identifies 
$ E_x = {g(r)}{r_x}/r $ and
$ E_y = {g(r)}{r_y}/r $
as the two components of the restoring force in \reff{rsm_eomCPL}. 
Note that
we have divided the electric field components by a
factor $\sqrt{2}$ so that the ponderomotive energy $U_\mathrm{p} =
\Ehat^2/4\omegalaser^2$, (i.e., the time-averaged quiver energy
of a free electron in the laser field)
is the same as in the LP case with the same $\Ehat$ (otherwise $U_\mathrm{p}$ would be a factor of two higher in the CP case).
The square of the effective, time-dependent oscillator frequency in the CP laser
field can be written as 
\beq
\left[\frac{\omega_{\mathrm{eff}}(\tau)}{\omegalaser}\right]^2 
=\frac{r_x E_x  + r_y E_x }{r^2} 
= \frac{g[r(\tau)]}{r(\tau)}
\label{efffrequCPL}, \eeq 
which has the same  form as in the
LP case \reff{efffrequ}.
%
%
%
\begin{figure}
\includegraphics[width=0.5\textwidth]{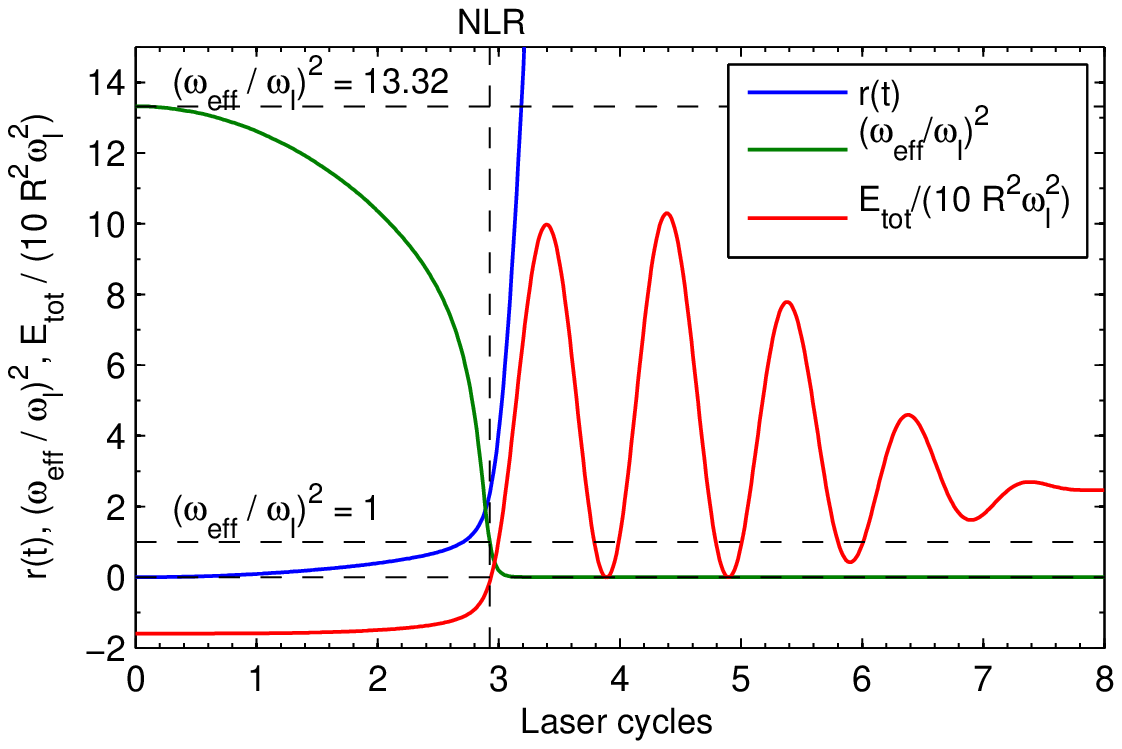}
\caption{(color online). Typical behavior of $\left({\omega_\mathrm{eff}(\tau)}/{\omegalaser}\right)^2$
(green, top left solid line) vs laser cycles above the threshold intensity
$\simeq 3\times 10^{16} \mathrm{W/cm^2}$ for circular polarization, $(\omegaMie/\omegalaser)^2=40/3$, 
and cluster radius $R = 3.2\,$nm.
Here, $\Ehat/R\omegalaser^2\simeq 8.2$, 
corresponding to a laser intensity
$\simeq 5\times 10^{16} \mathrm{W/cm^2}$,
$n=8$-cycle $\sin^2$-pulse with components
$E_\mathrm{l}^x(\tau)=\Ehat\sin^2(\tau/2n)\cos(\tau)/\sqrt{2}$,
$E_\mathrm{l}^y(\tau)=\Ehat\sin^2(\tau/2n)\sin(\tau)/\sqrt{2}$, and
wavelength $\lambda = 1056\,$nm. 
Excursion $r$ (blue, middle left solid line) and  energy
$\Etot/R^2\omegalaser^2$ (red, bottom left solid line) of the electron sphere 
are also plotted. The 
energy $\Etot/R^2\omegalaser^2$ is scaled down by a factor 10
to display within the excursion and the frequency range.
Outer ionization 
(i.e., $\Etot/R^2\omegalaser^2\geq 0$) and occurrence of
NLR 
$ \left[\omega_\mathrm{eff}(\tau)/\omegalaser\right]^2=1$
always coincide (dashed vertical). \label{rsmIICPL}}
\end{figure}
Earlier, in Fig.~\ref{rsmII}, it was shown that NLR and
outer ionization in the RSM only occur when a threshold laser intensity is
crossed. The same is true for the occurrence of NLR with
CP light.
Figure \ref{rsmIICPL} shows
 the temporal behavior of
$\left[\omega_\mathrm{eff}(\tau)/\omegalaser\right]^2$ 
 above the
threshold driver strength for a $n=8$-cycle CP laser
pulse of wavelength $\lambda = 1056\,$nm. Here the cluster charge 
density is $40$ times over-critical, i.e., 
$(\omegaMie/\omegalaser)^2=13.32$ at which 
$\left(\omega_\mathrm{eff}(\tau)/\omegalaser\right)^2$ starts 
and drops with increasing 
driver field during the pulse. The NLR $
\left[\omega_\mathrm{eff}(\tau)/\omegalaser\right]^2=1$ is passed at
the time indicated by the vertical line. As in the LP case in Fig.~\ref{rsmII}, the electron sphere is set free at the time the NLR is passed:  the energy of the electron sphere passes through zero, and the excursion 
sharply
increases to a high value. It is also clear from 
Fig.~\ref{rsmIICPL}, that once the electronic sphere is set free, the frequency drops to zero, and the total
absorbed energy remains positive. 
A zero effective frequency implies an infinite period, i.e, the 
electron sphere does not return to the ion sphere. 
The main difference between Fig.~\ref{rsmIICPL}
and Fig.~\ref{rsmII} is that in the
case of the CP laser field the decrease of the effective
frequency is smooth (i.e., no oscillations) since the electric field vector rotates but its absolute value remains constant. As a consequence, the electron sphere spirals out, staying away from the potential center where $\omega_\mathrm{eff} = \omegaMie$. For LP instead, the electron sphere is driven through the origin and hence the effective
frequency undergoes oscillations before it drops to the resonance
value, as visible in Fig.~\ref{rsmII}. NLR is clearly identified in both cases.

\subsection{Prediction of threshold intensity for the NLR}\label{sec2c}
NLR occurs above a threshold
driver strength. Beyond this driver strength the rigid 
electron sphere gains laser energy which is 
many order of magnitude higher than below the threshold (see Fig.~\ref{rsmI}).
In an open potential such as \reff{rsm_pot} the electron sphere is detached from the ion sphere above the threshold driver strength, i.e., appreciable energy absorption and outer ionization occur simultaneously. We would like to remark that in closed potentials NLR occurs as well, as discussed in Ref.~\cite{baumu05}.

The threshold driver strength can be estimated.
The dimensionless potential 
    $U(r) = {V(r)}/({\omegaMie^2 R^2 })$ 
of the electron sphere 
in the ionic field can be written as
\begin{equation}
    U(r) = \frac{r^2}{2} - \frac{3 r^3}{16}
    + \frac{r^5}{160}, \,\,\,\, r \le 2. \label{rsm_pot2}
\end{equation}
Application of a static
electric field $E_0$ (corresponding to the peak field strength of a low-frequency laser
field), suppresses the potential in one direction by the amount $R E_0\, r$. 
The effective potential seen by the electron sphere is
$U_\mathrm{eff} (r) =  U(r) - \hat E_0\, r$ with $\hat E_0 = E_0/(\omegaMie^2 R)$. 
The potential barrier vanishes if
$ U'(r_\mathrm{v}) - \hat E_0 = 0 $
and $    U''(r_\mathrm{v}) = 0$, leading to
$r_\mathrm{v} = 1$,  and the NLR threshold intensity is estimated to be
\begin{figure}
\includegraphics[width=0.5\textwidth]{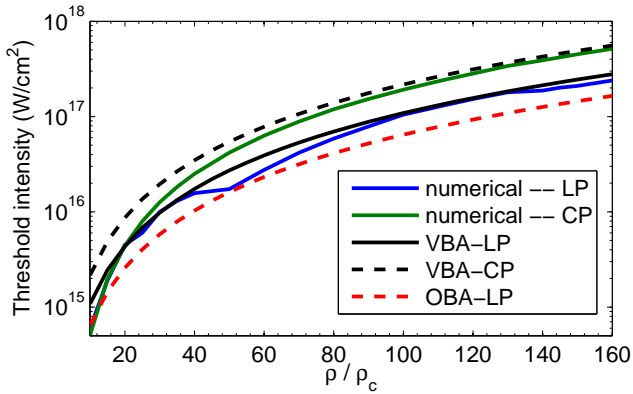}
    \caption{NLR threshold intensities in the RSM  vs the cluster
    charge density $\rho/\rhocrit$ for linearly (LP) and circularly
    (CP) polarized laser fields. 
    Results from the full numerical solution of 
    \reff{rsm_eom} with LP (lower solid),
    \reff{rsm_eomCPL} CP (top solid),
    the vanishing barrier approximation (VBA-LP, middle solid) 
    \reff{barappx}, and the VBA corrected for CP (upper dashed)
    are shown (see text for a discussion). 
    The over-the-barrier approximation (OBA) \reff{obaappx}
    (lower dashed)
    underestimates the exact threshold intensities.
    \label{rsmpert2}}
\end{figure}
\begin{equation}
I_\mathrm{th}^\mathrm{VBA}  
= E_0^2 = \left(\frac{5}{32}\frac{\rho}{\rhocrit} \omegalaser^2 R\right)^2.
\label{barappx}
\end{equation}
We call this the vanishing-barrier-approximation (VBA).

In atomic ionization the so-called over-the-barrier approximation (OBA) or Bethe-rule \cite{bethe} allows to estimate at which electric field strength a certain atomic charge state dominates. If one applies the OBA  to the RSM one obtains the two equations $U_\mathrm{eff} (r_\mathrm{b}) =  0$ and $ U'(r_\mathrm{b}) - \hat E_0 = 0$ with $r_\mathrm{b}$ the barrier location. One finds $r_\mathrm{b} \simeq 1.613$. This gives the OBA  
threshold intensity of the NLR
\begin{equation}
I_\mathrm{th}^\mathrm{OBA}  
 \simeq \left(\frac{10}{83}\frac{\rho}{\rhocrit} \omegalaser^2 R\right)^2,
\label{obaappx}
\end{equation}
which underestimates  
the numerically determined  threshold intensity, as we will show now.

Figure~\ref{rsmpert2} shows the threshold intensity as a function of the cluster
charge density $\rho/\rhocrit = 10$--$160$ (corresponding 
to average charge states $\simeq 1-16$ for Xenon). The numerically determined
threshold intensities for LP \reff{rsm_eom} and CP \reff{rsm_eomCPL}
laser light 
show that when the cluster charge density is low, the NLR occurs almost at the
same value of the threshold intensity, irrespective of the polarization. 
As the charge density increases the NLR threshold intensity appears to be higher for CP than for LP.
The VBA 
\reff{barappx} of the threshold intensity
is in good agreement with the numerical result for LP 
whereas the OBA \reff{obaappx}  underestimates it. This fact might be related to the recently observed ``enhanced saturation intensities'' in the ionization of finite size systems such as C$_{60}$ (see, e.g., \cite{hertel,JaronBecker} and references therein), indicating that the latter might neither be a many-electron nor a quantum effect but just due to the finite size of the target.

The difference of the threshold intensities for LP and CP is due to the definition of the CP field (\ref{Ex}), (\ref{Ey}) where a factor $2^{-1/2}$ has been introduced in order to render the ponderomotive potential equal for LP and CP. However, for the threshold intensity it is the electric field (or the intensity) that matters, not $\Up$. For a given $E_0$ the laser intensity is $I_0=E_0^2$ in the LP case but only $I_0/2$ for CP. 
Therefore, the upper black, dashed line 
in Fig.~\ref{rsmpert2} shows the VBA threshold intensity multiplied by a factor of two, which is in good agreement with the numerical results for the CP laser field at higher charge densities.

So far we have studied the NLR absorption of laser
energy in a simplified model system assuming an anharmonic 
potential generated by the ions in which the homogeneous and rigid
electron cloud moves.
In reality, the potential builds up during the interaction with 
the laser pulse because of ionization. The delicate interplay of inner ionization, energy absorption by various mechanisms, and outer ionization can be simulated using methods such as PIC or molecular dynamics. Previous work \cite{kundu06}, studying 
LP short laser pulses,
showed that NLR
\reff{nlr} can be clearly identified in such simulations 
as well. 
\section{Nonlinear resonance: particle-in-cell results}\label{sec3}
In this Section we present results obtained from three-dimensional PIC simulations.
The cluster is exposed to $n=8$-cycle $\sin^2$-pulses
$E_\mathrm{l}(t)=\Ehat\sin^2(\omegalaser t/2n)\cos(\omegalaser t)$
of near infrared wavelength $\lambda=1056$\,nm, i.e.,
the total pulse duration is $28\;\mathrm{fs}$.
Since motion of ions does not play an
important role during the entire pulse the ions are assumed fixed, which
ensures a well defined, constant Mie frequency $\omegaMie$.
A PIC electron has the same charge to mass
ratio as a ``real'' electron, that is, $e/m=-1$ in atomic units.
Each PIC electron moves under the influence of the external laser
field and the space charge field  $\vektEsc=-\vektnabla
\Phi(\vektr,t)$ due to the potential $\Phi(\vektr,t)$ that is
created by all charges (mapped to the numerical grid).
Hence the
equation of motion of the $i$-th PIC electron is
\beq \ddot{\vektr}_i
+ \vektEsc(\vektr_i,t) = -\vektE_\mathrm{l}(t).\label{eqpic}
\eeq
Equation \reff{eqpic} is solved self-consistently for all PIC electrons.
Clearly, $\vektEsc(\vektr_i,t)$ depends on
the position of {\em all} other particles $\neq i$ as well. The PIC
simulation starts with the neutral cluster configuration so that $\vektEsc(\vektr_i,0)\equiv
\vekt{0}$. 
\subsection{Results for linear polarization}
\begin{figure}
\includegraphics[width=0.5\textwidth]{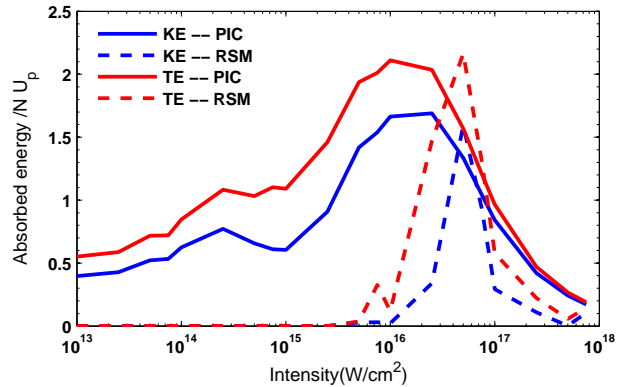}
\caption{(color online). Kinetic energy (KE) and total absorbed (TE)
energy per electron in units of $\Up$ vs  laser intensity for charge
density $40$ times the critical density and a 
Xe$_\mathrm{1611}$ cluster ($\mathrm{N}=5498$ electrons, 
mean ionic charge state $\simeq 3.4$, and a cluster radius $R = 3.2\,$nm).
Solid lines are PIC results,
dashed lines are the RSM results for
a $n=8$ cycle pulse
$E_\mathrm{l}(t)=\Ehat\sin^2(\omegalaser t/2n)\cos(\omegalaser t)$ of wavelength
$\lambda = 1056 $~nm.
\label{picrsm1}}
\end{figure}
Figure \ref{picrsm1} shows PIC
results  for the kinetic energy (KE) and the total energy (TE)
absorbed per electron in units of 
$\Up$, vs the peak laser
intensity.
One sees that the absorbed energy per electron is on the order of $\Up$.
However, the absorbed energy is nonlinear in $\Up$ and displays a
maximum before it drops because
of the saturation of outer ionization. 
The depletion is due to the fact that
at a given intensity most (if not all)  electrons are removed from the cluster (complete outer ionization). Further inner ionization would be required to generate ``fresh'' electrons that could continue to absorb energy.
The maxima in the PIC absorption curves 
are located close to the 
threshold intensity predicted by the RSM.
The total absorbed energy around the
maximum of the PIC absorption curves (RSM as well)
is on the order of $2 U_\mathrm{p} = 5$--$6 $\,keV which has been
also reported in experiments 
of intense laser clusters
interactions \cite{kum03}. 

In PIC simulations no sharp
intensity threshold exists since each PIC electron sees its own
time-dependent field (space charge field plus the laser field).
Therefore sharp jumps (as seen in the RSM, e.g., Fig.~\ref{rsmI})
are absent.

\begin{figure}
\includegraphics[width=0.5\textwidth]{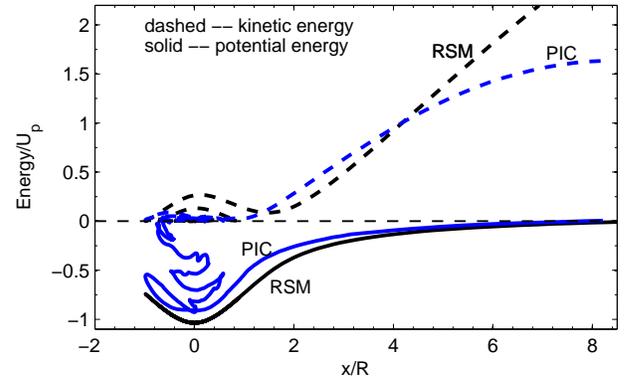}
    \caption{(color online). Comparison between the kinetic and the
    potential energies of a 
PIC electron and the equivalent 
RSM electronic particle with same charge 
and mass of a PIC electron vs the excursion 
$x/R$ in the laser polarization direction. 
The escaping PIC electron resembles the equivalent 
electronic particle
in the RSM. 
Peak laser intensity $I_0 = 2.5\times 10^{16}$ W/cm$^2$,  charge
density is $\rho/\rhocrit=40$. Other parameters as in Fig.~\ref{picrsm1}.
\label{picrsm2}}
\end{figure}

Figure \ref{picrsm2} compares the kinetic
and potential energies of a single PIC electron and
the electronic sphere in the RSM. The energies are plotted vs the excursion $x/R$ (along
the direction of the laser field).  The peak laser intensity is $I_0 = 2.5\times 10^{16}$ W/cm$^2$.
We have checked that there are many PIC electrons
that leave the cluster similarly to the one presented in
Fig.~\ref{picrsm2}.
The potential and the  
kinetic energy gained by the electron sphere in the RSM 
is divided by the total number of 
PIC electrons so that a quantitative comparison is possible. 
The main difference between PIC and RSM results is that in the PIC  simulation all the electrons see initially a zero potential since the potential builds up from the neutral cluster configuration in the course of the interaction with the laser field while in the RSM the electronic sphere oscillates
in a prescribed potential. Those PIC electrons that stay long inside the cluster experience almost the full ionic potential which then is similar to the RSM potential. 

As mentioned earlier, the absorption of energy by a PIC electron
depends upon the self-consistent potential
which
develops during the laser pulse due to outer ionization.
As a result different PIC electrons move along different trajectories, ``see'' a different potential, and thus are set free at different times.
\begin{figure}
\includegraphics[width=0.5\textwidth]{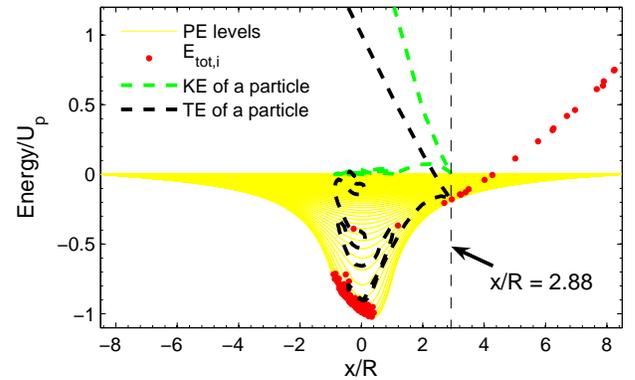}
    \caption{(color online). Energy and self-consistent potential $\Phi$
from the PIC simulation vs the excursion $x/R$
in laser polarization direction.
Yellow (solid gray) lines represent cuts of the potential $\Phi$  at time
$\omegalaser t/2\pi = 4.81$ for different $y$ and $z=0$.
The circles (red) represent the total energy $E_{\mathrm{tot},i}$ of
individual PIC electrons located within the simulation box at that time.
The dashed lines represent the kinetic (green, dashed gray) energy (KE)
and potential (dashed black) energy (PE) of a
PIC electron that is outer ionized  when the excursion $x/R \simeq 2.88$ meets the condition
$\omega_\mathrm{eff} = \omegalaser$ at the same time (see Fig.~\ref{rsmIII} for the charge density
$\rho/\rhocrit = 40$).
The peak laser intensity is $I_0 = 2.5\times 10^{16}$ W/cm$^2$.
Other parameters as in Fig.~\ref{picrsm2}.
\label{picpot}}
\end{figure}
Figure \ref{picpot} shows a
snapshot of the collective potential 
$\Phi(\vektr,t)$ at time
$\omegalaser t/2\pi = 4.81$ for a cut at $z=0$  (and various $y$ throughout the cluster). 
The
most lower curve represents the potential for $y=z=0$.
The red circles  represent the total energy 
$E_{\mathrm{tot}, i} = \dot{\vektr}_i^2(t)/2-\Phi(\vektr_i,t)$ 
of individual PIC electrons located within the simulation box at that time.

One observes that many PIC electrons are accumulated
near the bottom-left of the potential well 
at this time. These PIC electrons with $E_{\mathrm{tot}, i}<0$ remain bound  since the laser field amplitude is already decreasing 
from the 4th cycle onward. The kinetic  and 
the total energy 
of a PIC electron which leaves
the potential well when its excursion becomes $x/R \simeq 2.88$ at the same time is also shown. This excursion
approximately satisfies the NLR condition
$\omega_\mathrm{eff}/\omegalaser = 1$ for the charge density $\rho/\rho_\mathrm{c} = 40$,
as identified in Fig.~\ref{rsmIII} with the RSM analysis.

The NLR behavior exhibited by
the PIC electron in
Fig.~\ref{picpot} is not accidental.
For the sake of an unequivocal and explicit identification of the NLR we now analyze
the motion of all individual PIC electrons
in the same way as it has been done with the motion of the electron
sphere in the RSM in Sec.~\ref{sec2}.
Recalling \reff{eqpic}, 
the equation for the
effective, time-dependent oscillator frequency analogous to
\reff{efffrequCPL} for the $i$-th PIC electron reads
\beq \omega^2_{\mathrm{eff},i}(t) =
-\frac{[\vektE_\mathrm{l}(t)+\ddot{\vektr}_i(t)]\cdot\vektr_i(t)}{\vektr^2_i(t)}=\frac{\vektEsc(\vektr_i,t)\cdot\vektr_i(t)}{\vektr^2_i(t)}
\label{efffrequPIC}.
\eeq
Earlier we have mentioned that $\vektEsc(\vektr_i,t)$ depends on
the position of {\em all} other particles $\neq i$ and the 
simulation starts with the charge neutral cluster configuration 
i.e., $\vektEsc(\vektr_i,0)\equiv
\vekt{0}$. Hence, a PIC electron  ``sees'' initially an effective
frequency $\omega_{\mathrm{eff},i}(0)=0$. The laser field disturbs
the charge equilibrium and $\omega^2_{\mathrm{eff},i}(t)$ becomes
different from zero. $\omega^2_{\mathrm{eff},i}(t)$ may be even
negative in regions of accumulated electron density (repulsive
potential). As the cluster charges up,
$(\omega_{\mathrm{eff}}/\omegalaser)^2$ quickly increases beyond
unity (where the RSM starts in the first place).  The start from
$\omega_{\mathrm{eff},i}(0)=0$, the possibility of negative
$\omega^2_{\mathrm{eff},i}(t)$, and the three-dimensionality are the
main differences to the RSM analysis above.
\begin{figure}
\includegraphics[width=0.5\textwidth]{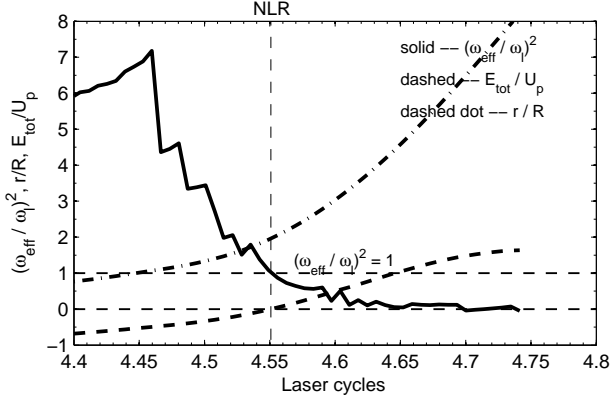}
\caption{Effective frequency squared ($\omega_{\mathrm{eff}, i}/\omegalaser)^2$,
    excursion $r_i/R$,
    and the total energy
    $E_{\mathrm{tot}, i} = \dot{\vektr}_i^2(t)/2-\Phi(\vektr_i,t)$ for
    the PIC electron of Fig.~\ref{picrsm2} vs time in laser
    cycles.
    The total energy becomes positive only when
the NLR is crossed (indicated by the vertical dashed line).
This result resembles the RSM result in
Fig.~\ref{rsmII} when NLR is met.
The charge density is $\rho/\rhocrit = 40$, the peak laser intensity is $I_0 = 2.5\times 10^{16}$ W/cm$^2$.
Other parameters as in  Fig.~\ref{picrsm2}.
    \label{timeVsOmega}
 }
\end{figure}
Figure \ref{timeVsOmega}
shows the effective frequency squared, the total energy
$E_{\mathrm{tot}, i}(t)$, and the
excursion $r_i/R$ vs time for the PIC electron
whose energy history is shown
in Fig.~\ref{picrsm2}. 
We define the time when, for a particular electron, $E_{\mathrm{tot}, i}$
becomes $>0$ as the ionization time of
that electron.
Data are plotted shortly
before the emission of this particle
from the cluster potential.
It is clearly visible in Fig.~\ref{timeVsOmega}
that the PIC electron is escaped
only when the resonance line
$(\omega_{\mathrm{eff}}/\omegalaser)^2 = 1$ is passed. 
Figure~\ref{timeVsOmega} can be well compared with 
Fig.~\ref{rsmII} showing ionization of the RSM via NLR.

In the
case of PIC simulations the fulfillment of the nonlinear 
resonance condition
$(\omega_{\mathrm{eff}}/\omegalaser)^2 = 1$ is necessary but 
not sufficient for ionization. As the potential builds up, the PIC electrons transiently meet the NLR condition, and, in fact, some electrons leave the cluster at that early stage when the potential is still shallow and the laser field is relatively weak. However, as the potential deepens, PIC electrons ``dropping'' below the energy necessary for NLR to occur, behave from then on similar to the RSM and may finally escape only by climbing up in the potential and hitting the NLR $(\omega_{\mathrm{eff}}/\omegalaser)^2
= 1$.
During the 0.4 laser cycles plotted in Fig.~\ref{timeVsOmega} the $(\omega_{\mathrm{eff}}/\omegalaser)^2$-curve displays artificial, short-time scale fluctuations inherent in PIC simulations \cite{birdsall}. However, we checked that macroscopic observables such as the absorbed energy or the degree of outer ionization are well converged.

\begin{figure}
\centering\includegraphics[width=0.55\textwidth]{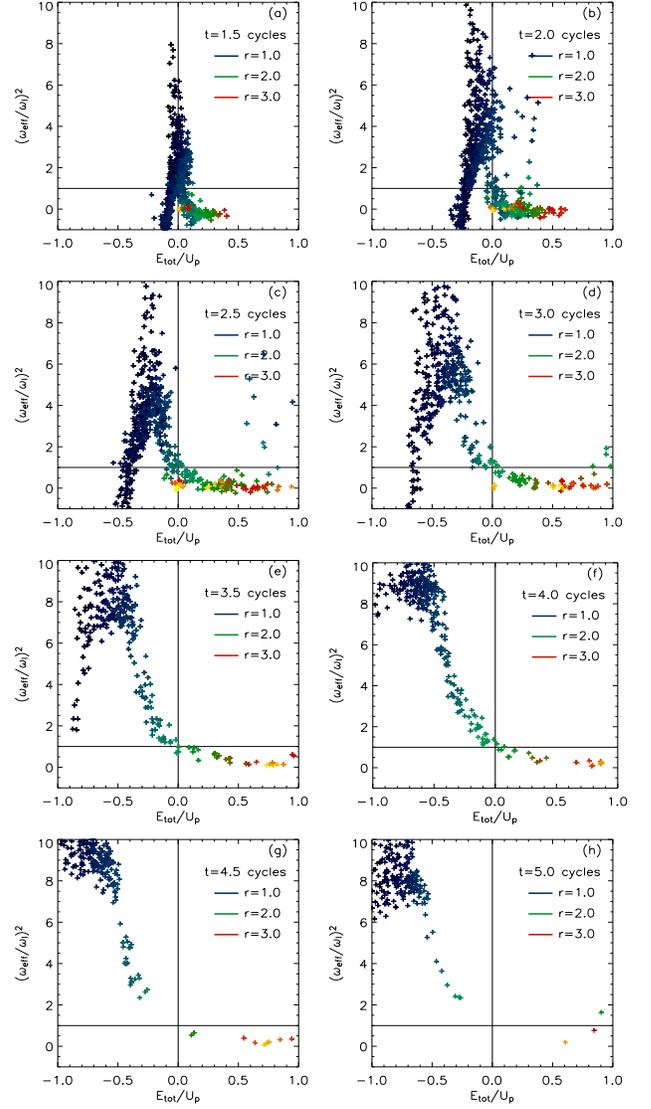}
\caption{(color online). Snapshots of PIC electrons in the frequency
vs  energy-plane at times
(a) $t=1.5$, (b) $t=2.0$, (c) $t=2.5$, (d) $t=3.0$,
(e) $t=3.5$, (f) $t=4.0$, (g) $t=4.5$, and (h) $t=5.0$ laser cycles
for LP,
laser intensity $2.5\times
10^{16}$\,\Wcmcm, and $(\omegaMie/\omegalaser)^2=40/3$. Other
parameters as in Fig.~\ref{timeVsOmega}.
The radial positions (in
units of $R$) are color-coded. Electrons become free upon crossing
the NLR, i.e.,
$(\omega^2_{\mathrm{eff}}/\omegalaser^2,\Etot/\Up)=(1,0)$.
\label{PICweffvsEngLP}}
\end{figure}

By following the dynamics of the electrons in the effective
frequency vs  energy-plane we identify the main pathway to outer
ionization and efficient absorption. Figure~\ref{PICweffvsEngLP}a--h
shows the scaled effective frequencies squared
$(\omega_{\mathrm{eff}}/\omegalaser)^2$ of the individual PIC
electrons vs  their energies
$\Etot(t)=\dot{\vektr}_i^2(t)/2-\Phi(\vektr_i,t)$ they would have if
the driver is switched off instantaneously at $t=1.5$, $2$, $2.5$,
$3$, $3.5$, $4$, $4.5$, and $5$ laser cycles, respectively. At the
time when, for a particular electron, $\Etot$ becomes $>0$  the
ionization occurs for that electron. The laser intensity is
$2.5\times 10^{16}$\,\Wcmcm, and the pre-ionized cluster is $40$
times over-critical so that $(\omegaMie/\omegalaser)^2=40/3$. As is
clearly visible in Fig.~\ref{PICweffvsEngLP}, each electron reaches
positive energy close to the point
$(\omega^2_{\mathrm{eff}}/\omegalaser^2,\Etot/\Up)=(1,0)$. The
radial position of each electron is color-coded, indicating that
outer ionization occurs at radii around $2R$. During the early time of
the laser pulse (Fig.~\ref{PICweffvsEngLP}a,b) when many electrons are still inside the
cluster, $(\omega_{\mathrm{eff}}/\omegalaser)^2$ spreads over
a wide range, starting from the maximum value
$(\omegaMie/\omegalaser)^2$ down to negative values due to the
repulsive force exerted by the compressed electronic cloud. Note
that negative values in effective frequency occur mainly at early
times where most of the electrons are still inside the cluster.
Electrons with positive but very
small $\Etot$ and $\omega^2_{\mathrm{eff}}\simeq 0$ represent low
energetic electrons removed earlier during the pulse (see Fig.~\ref{PICweffvsEngLP}a,b). The occurrence
of NLR is less clear for these early leaving electrons. As mentioned
above, these electrons move in a shallow effective potential with
$(\omega_{\mathrm{eff}}/\omegalaser)^2<1$ when they leave the
cluster with ease and with rather low kinetic energy because the
laser intensity is still low at the time of their emission.
Figures~\ref{PICweffvsEngLP}c--f show that most of the electrons escape from the cluster by passing through the channel
$(\omega^2_{\mathrm{eff}}/\omegalaser^2,\Etot/\Up)=(1,0)$ at radii
around $2R$. It is also visible that more and more electrons are driven
to positive frequency before they leave the cluster by passing through
$(\omega^2_{\mathrm{eff}}/\omegalaser^2,\Etot/\Up)=(1,0)$. This is
so because as more and more electrons are freed, the
remaining electrons experience predominantly the force by the ionic
background, and they move deep into the potential (see their
negative values in energy) where they experience the full Mie-frequency
$\omegaMie/\omegalaser = 40/3$. In Fig.~\ref{PICweffvsEngLP}c,d,
the few electrons with positive energy but small radii  are those driven back to the
cluster by the laser field. In Figures~\ref{PICweffvsEngLP}e,f,
electrons are strongly aligned (no scattered points) since the laser
field is approaching its maximum (at $t = 4$-cycles). After the
peak of the laser pulse (Figs~\ref{PICweffvsEngLP}g,h) the
restoring force of the ions on almost all electrons dominates the laser
force.

\subsection{Results for circular polarization }
Since many of the features of energy absorption and NLR
in a LP laser field are common to the case of
CP, we only point out the main differences.
\begin{figure}
\includegraphics[width=0.5\textwidth]{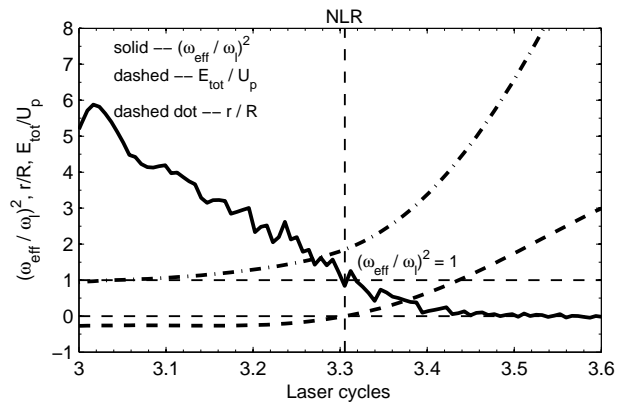}
    \caption{Effective frequency squared 
    ($\omega_{\mathrm{eff}, i}/\omegalaser)^2$,
    excursion $r_i/R$,
    and the total energy
    $E_{\mathrm{tot}, i} = \dot{\vektr}_i^2(t)/2-\Phi(\vektr_i,t)$
    for a PIC electron in a CP field vs time in laser cycles.
    The total energy  
    becomes positive only when
the NLR is crossed (indicated by vertical, dashed line).
This result resembles the RSM result in 
Fig.~\ref{rsmIICPL}. The charge density is $\rho/\rhocrit = 40$,
the peak laser intensity is $I_0 = 2.5\times 10^{16}$ W/cm$^2$.
    \label{timeVsOmegaCPL}
 }
\end{figure}
Equation \reff{efffrequPIC} holds in the CP
laser field as well.
Figure \ref{timeVsOmegaCPL} shows the effective
frequency squared vs time for one of the PIC electrons, 
together with the total energy
$E_{\mathrm{tot}, i}(t)$ and
the excursion $r_i/R$. One can see that the PIC electron is freed (i.e., its total
energy becomes positive) only when the resonance line
$(\omega_{\mathrm{eff}}/\omegalaser)^2 = 1$ is passed.

\begin{figure}
\includegraphics[width=0.55\textwidth]{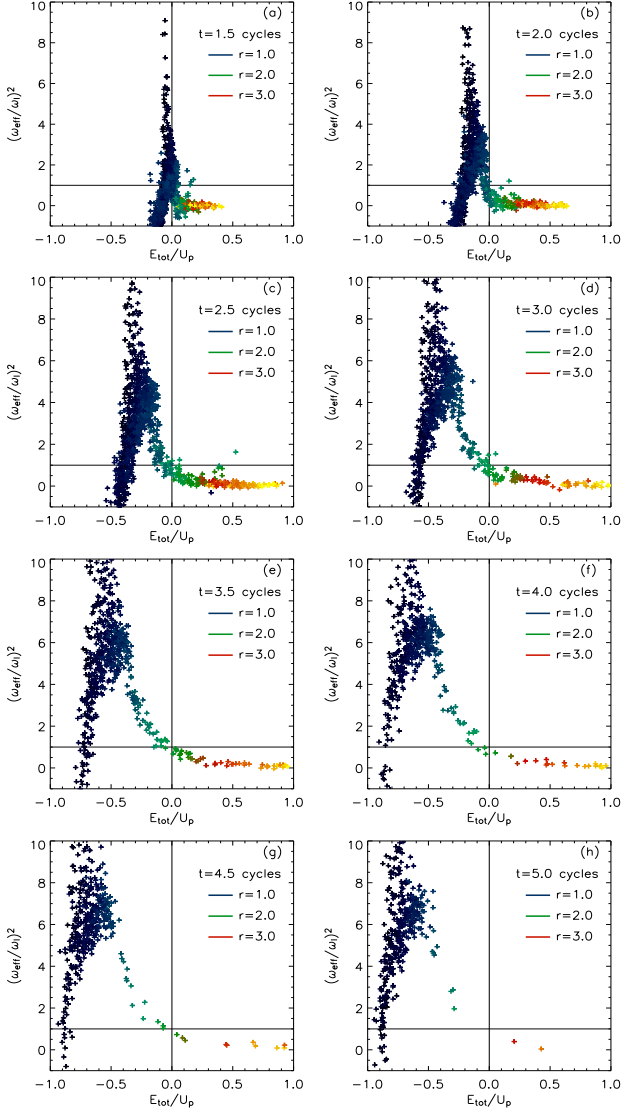}
\caption{(color online). Snapshots of PIC electrons in the frequency
vs energy-plane for CP at times
(a) $t=1.5$, (b) $t=2.0$, (c) $t=2.5$, (d) $t=3.0$,
(e) $t=3.5$, (f) $t=4.0$, (g) $t=4.5$, and (h) $t=5.0$ laser cycles for
a laser intensity $2.5\times
10^{16}$\,\Wcmcm and $(\omegaMie/\omegalaser)^2=40/3$. Other
parameters as in Fig.~\ref{picrsm1}. The radial positions (in
units of $R$) are color-coded. Electrons become free upon crossing
the NLR, i.e.,
$(\omega^2_{\mathrm{eff}}/\omegalaser^2,\Etot/\Up)=(1,0)$.
\label{PICweffvsEngCPL}}
\end{figure}

Figure~\ref{PICweffvsEngCPL} is the CP analogue of Fig.~\ref{PICweffvsEngLP}.
The results are very much similar to the LP case
shown in Fig.~\ref{PICweffvsEngLP} and the arguments made
there apply here as well. NLR is clearly observed.
The main difference is 
that the PIC electrons are nicer aligned towards the resonance point,
even at early times during the laser pulse (see
Fig.~\ref{PICweffvsEngCPL} b, c, d). Almost no scattered particles are
visible because the dynamics mainly consist of swirling around the cluster center rather than oscillating through it. The number of electrons returning to the cluster is much less so that the recombination and rescattering 
probability is smaller in the case of CP. The same is observed in laser-atom interaction experiments, with important consequences for harmonic generation and non-sequential ionization.

For sufficiently high charge 
density, the degree of outer ionization (defined
as the ratio of removed electrons to the total number of
electrons) is less compared to the case of LP  laser pulses. The explanation is the same as in Sec.\ref{sec2c}: the effective laser intensity for CP  is by a factor of two lower than for LP because of our definition of the CP field (\ref{Ex}), (\ref{Ey}).

One may object that, since the denominator in \reff{efffrequPIC}
necessarily increases while the numerator decreases for an electron
on its way out of the cluster potential, that the passage through a point
$(\omega^2_{\mathrm{eff}}/\omegalaser^2,\Etot/\Up)=(x,0)$ with $x$
{\em some} value $< (\omegaMie/\omegalaser)^2$ is rather the
consequence of outer ionization than the mechanism behind it.
However, NLR only occurs at $x=1$, and the results in
Fig.~\ref{PICweffvsEngLP} and
Fig.~\ref{PICweffvsEngCPL} 
show only  little spreading along
$(\omega_{\mathrm{eff}}/\omegalaser)^2$ at $\Etot=0$. Moreover, the
fact that {\em both} the single electron energies become positive
{\em and}  the radii exceed $\simeq 2 R$ when
$(\omega_{\mathrm{eff}}/\omegalaser)^2= 1$ indicates that NLR is
indeed the responsible mechanism behind outer ionization accompanied
by efficient absorption of laser energy.
%
%
\begin{figure}
\includegraphics[width=0.48\textwidth]{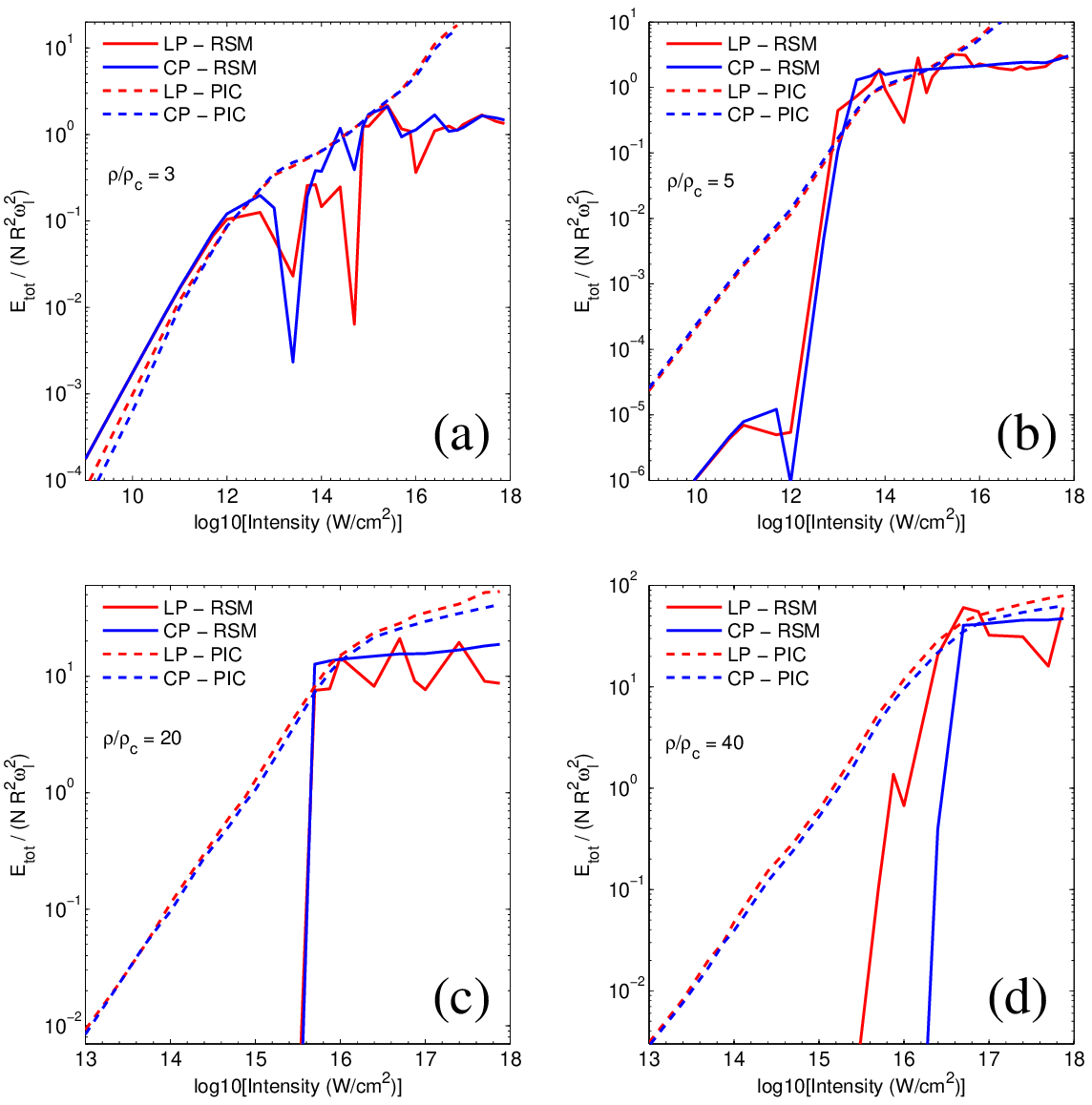}\\
\includegraphics[width=0.48\textwidth]{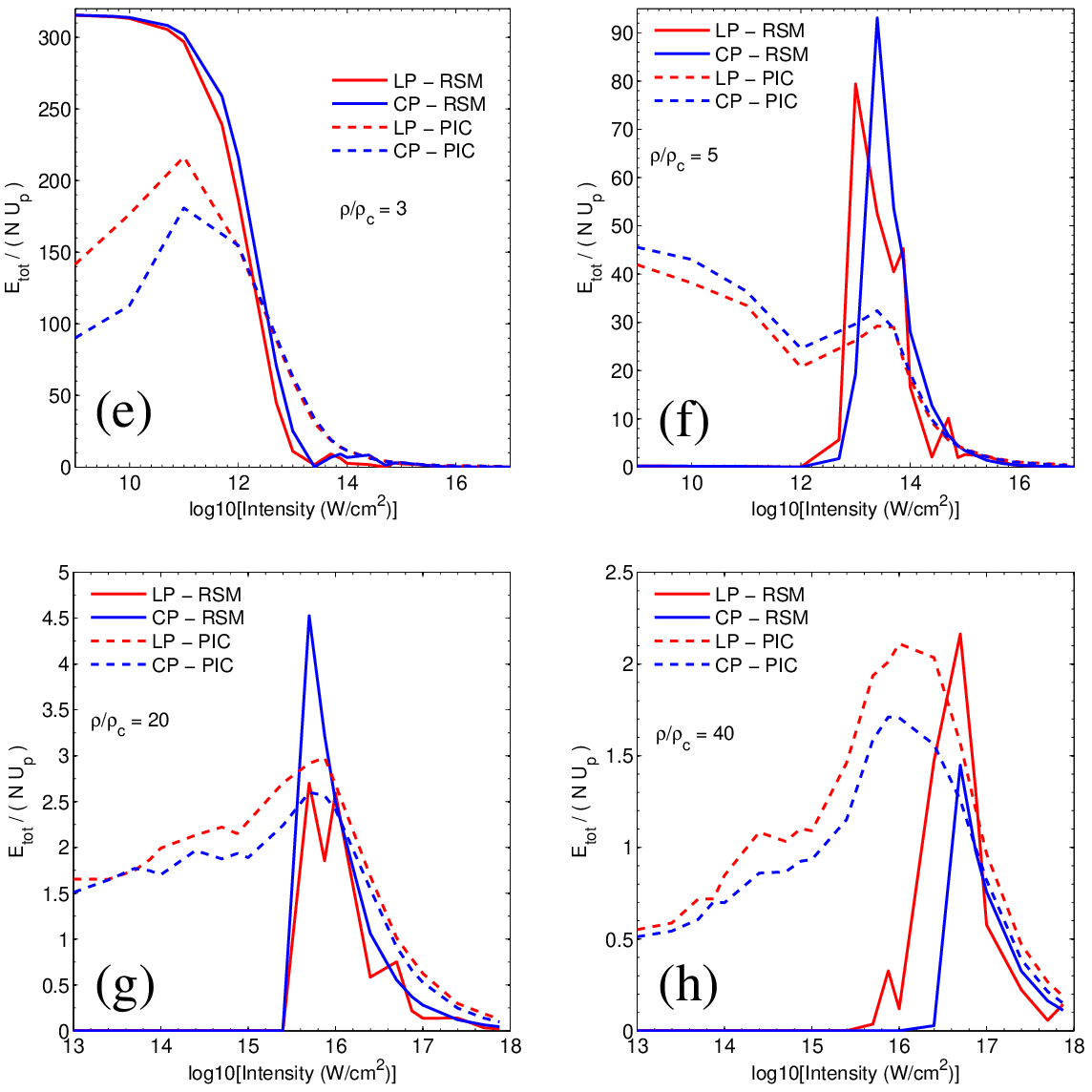}
\caption{(color online). Total absorbed
energy per electron in units of $R^2\omegalasersq$ vs  laser intensity for charge
densities 
(a) $\rhorat = 3$ (linear resonance), 
(b) $\rhorat = 5$, 
(c) $\rhorat = 20$, 
and (d) $\rhorat = 40$. 
The absorbed energies
 for LP (red, lighter gray) and CP (blue, darker gray) using PIC (dashed)
and the RSM (solid) are shown.
\label{rsmEngCPL}}
\end{figure}

Finally, Fig.~\ref{rsmEngCPL} shows the average value of the
total absorbed energy per electron vs the peak laser intensity
for cluster charge densities between $\rho/\rhocrit = 3$--$40$ 
in CP  and LP  laser fields. 
PIC results  are compared with the RSM 
absorption results.
The absorbed energy per electron in Figs.~\ref{rsmEngCPL}a--d is plotted 
in units of $R^2\omegalasersq$ whereas the same results are shown 
in Figs.~\ref{rsmEngCPL}e--h 
in units of the ponderomotive energy $U_\mathrm{p}$. 
The PIC results in Figs.~\ref{rsmEngCPL}a--d show that the absorbed energy increases linearly in the log-log representation up to a certain intensity and then tends to saturate due to the saturation of outer ionization. 
One sees that the saturation in the PIC
results occur close to the RSM threshold intensity. 
Since with increasing  charge density the restoring force due to the ions increases, the saturation of energy absorption in the RSM and  PIC occur at higher laser intensities as the density increases
from $\rho/\rhocrit = 3$ to $\rho/\rhocrit = 40$ in Fig.~\ref{rsmEngCPL}a--d. 
When outer ionization and the energy absorption saturate with increasing peak laser intensity, 
the average absorbed energy per electron divided by 
$U_\mathrm{p}$ (which is proportional to the so-called fractional absorption) starts decreasing in Figs.~\ref{rsmEngCPL}e--f.  
Figures \ref{rsmEngCPL}a,e 
show that at linear resonance 
$\rhorat = 3$, the absorbed energy is already high at low values of the laser intensity ($<10^{12}\, \Wcmcm$), and absorption 
is very efficient as compared to higher charge densities
$\rhorat = 5$--$40$, presented in Figs.~\ref{rsmEngCPL}b--d
and Figs.~\ref{rsmEngCPL}f--h. 
In fact, Fig.~\ref{rsmEngCPL}e illustrates that the absorbed energy is on the
order of $ \sim 100 U_\mathrm{p}$ (both in the RSM and in the PIC) 
before the saturation of 
outer ionization. However, one should bear in mind that at too low laser intensities inner ionization would not occur in the first place so that in reality there would be no absorption at all.

Energy absorption in CP and 
LP laser fields at all 
intensities and all charge densities are almost equally efficient. Recalling that with our definition of the CP laser field the ponderomotive potential is equal for LP and CP while the electric field amplitude is not, we conclude that for the absorbed energy $\Up$ matters while the NLR threshold is determined by the field strength.

The discrepancy between the PIC and RSM results at lower intensities 
is, again, due to the  fact that the RSM provides a very well defined potential, although a more and more shallow one as the density decreases,  in which the rigid electron sphere moves. In PIC, the potential is initially zero and builds up in the course of outer ionization. 
At higher intensities, 
the potentials for the remaining PIC electrons are closer to the RSM 
potential (see Fig.~\ref{picrsm2}), explaining the improving agreement 
between RSM and PIC results as the laser intensity increases.

\section{Summary}\label{sec4}
In summary, two different
approaches to study collisionless laser energy absorption by clusters, namely (i) the rigid sphere model and (ii)
particle-in-cell simulations, were pursued in this work.
The goal was to identify the dominant mechanism
of energy absorption and outer ionization
of the cluster electrons in near infrared, short laser pulses where collisional absorption is known to be inefficient.
We showed that the cluster electrons contributing to
efficient absorption and outer ionization 
undergo nonlinear resonance, meaning that the instantaneous
frequency of their motion in a time-dependent, anharmonic, effective
potential transiently meets the laser frequency. Nonlinear resonance is the only
possible absorption mechanism if the laser pulse is too short for
the linear resonance to occur (or during the early cluster dynamics
in longer pulses) and if electron-ion collisions (inverse
bremsstrahlung) are negligible. In order to prove the occurrence of
nonlinear resonance we used  a method to analyze the results
obtained from particle-in-cell simulations, namely the mapping of
the system of electrons and ions that interact through their mean
field  onto a system of nonlinear oscillators.

The occurrence of nonlinear resonance in the particle-in-cell
simulations presented in this paper resembles the
nonlinear resonance in the rigid sphere model. For a given cluster
charge density, there is a threshold intensity around which the average
electron energy displays a maximum conversion of laser energy. The
threshold intensity can be calculated using the newly introduced vanishing barrier
approximation. The common over-barrier approximation---applicable to atoms---fails in the case of finite-size potentials and underestimates the required laser field strength for ionization. This fact might be related to the experimentally observed ``enhanced saturation intensity'' in the ionization of finite-size systems such as C$_{60}$. 

The efficiency of energy absorption from the laser is almost the same for linear and circular polarization, indicating that ``collisions with the cluster boundary'' as an explanation are misleading. Instead, nonlinear resonance is the main absorption mechanism in both cases. 
To illustrate this, the rigid 
sphere model has been extended for the circularly polarized laser pulses 
in this paper.

Future work will take self-consistent charge state distributions and mobile ions into account. Preliminary results indicate that nonlinear resonance clearly persists under these circumstances as well.

This work was supported by the Deutsche Forschungsgemeinschaft.

\end{document}